\documentstyle[aps,prd]{revtex}
\tighten
\renewcommand{\epsilon}{\varepsilon}
\begin{document}
\title{Cosmological perturbation theory and conserved quantities 
in the large-scale
limit}

\author{Winfried Zimdahl}
\address{Fakult\"at f\"ur Physik, Universit\"at Konstanz, PF 5560 M678 
D-78434 Konstanz, Germany}

\maketitle

\begin{abstract}

The linear cosmological perturbation theory of an almost homogeneous and
isotropic perfect fluid universe is reconsidered and formally  
simplified by
introducing new covariant and gauge-invariant variables with physical
interpretations on hypersurfaces of constant expansion, constant  
curvature
or constant energy density. The existence of conserved perturbation
quantities on scales larger than the Hubble scale is discussed. The  
quantity
which is conserved on large scales in a flat background universe may be
expressed in terms of the fractional, spatial gradient of the  
energy density
on constant expansion hypersurfaces or, alternatively, with the help of
expansion or curvature perturbation variables on hypersurfaces of  
constant
energy density. For nonvanishing background curvature the perturbation
dynamics is most suitably described in terms of energy density  
perturbations
on hypersurfaces of constant curvature.

\end{abstract}


\section{Introduction}

Understanding the dynamics of small inhomogeneities in the early  
universe is
essential for any theory of cosmological structure formation. The
conventional, linear cosmological perturbation theory relies on a  
splitting
of all metric and matter quantities into a homogeneous and  
isotropic zeroth
order and small, first-order perturbations about this background. This
procedure implies a splitting of the spacetime itself. The  
nonuniqueness of
this splitting is the source of the so-called gauge problem \cite
{Bardeen,KoSa,EB,Stew}. A perturbation analysis may either be  
performed in a
specific gauge, the traditionally favoured gauge used to be the  
synchronous
gauge (see, e.g., \cite{Weinberg}), or in a gauge-invariant manner.  
While
the choice of a specific gauge corresponds to a specific  
identification of
the points of the fictitious background spacetime with those of the real
spacetime, a gauge-invariant description is independent of this
identification. Therefore, a gauge-invariant approach is  
conceptionally more
attractive although the physical meaning of the gauge-invariant  
variables in
terms of which the theory is formulated becomes obvious only for  
specific
observers.

According to their extension with respect to the Hubble scale,  
cosmological
perturbations are divided into large-scale perturbations, i.e.,
perturbations with wavelengths larger than the Hubble length, and
small-scale perturbations for which the opposite is true.

In inflationary cosmology the presently observed structures are  
traced back
to quantum fluctuations during an early de Sitter phase. These  
originally
small-scale perturbations are stretched out tremendously in the  
inflationary
period, thereby crossing the Hubble length (which is constant  
during the de
Sitter stage) and becoming large-scale perturbations. Lateron, after the
inflation has finished, i.e., when the universe is adequately  
described by
the standard Friedmann-Lema\^{\i}tre-Robertson-Walker (FLRW) model,  
these
perturbations cross the Hubble length again, this time inwards, and  
again
become small-scale perturbations.

Consequently, it is of interest to follow the perturbation dynamics  
during
the time interval between both Hubble scale crossings. Bardeen  
\cite{Bardeen}
, Bardeen, Steinhardt and Turner \cite{BST} were the first to  
recognize the
existence of a conserved quantity on large scales which simplified the
corresponding perturbation analysis. The existence of a conservation
quantity allows one to establish a link between physical variables  
in remote
cosmological periods. This well-known fact has been widely used in the
literature (see, e.g., \cite{Lyth,Jai,MFB,LiLy}).

Recently, the role of conserved quantities was discussed in terms of the
gauge-invariant and covariant Ellis-Bruni variables \cite{EB} by  
Dunsby and
Bruni \cite{DB}. The key quantities of the Ellis-Bruni analysis are the
covariantly defined spatial gradients of the energy density, the  
pressure,
the expansion and the 3-curvature scalar. From their definition all  
these
variables have a general physical meaning for a comoving (with the fluid
4-velocity) observer without any reference to perturbation theory.  
For small
deviations from homogeneity and isotropy they may be related to usual
perturbation variables which are gauge-invariant by construction. The
relations among the Ellis-Bruni variables in linear order 
and the corresponding results for the
cosmological modes coincide with those of the comoving gauge in  
conventional
perturbation theory \cite{EHB}. For different gauges, the Ellis-Bruni
quantities, if regarded as conventional, linear perturbation  
quantities (see
below) have no obvious physical meaning, in general. In other  
words, as to
their physical interpretation, comoving observers are preferred.

The present paper introduces a set of covariant and gauge-invariant
quantities which represent suitable combinations of Ellis-Bruni type
variables. Like the original Ellis-Bruni quantities, the new  
variables are
general tensorial expressions which may be used to characterize  
deviations
from homogeneity and isotropy without explicitly introducing a  
fictitious
background universe, i.e., they are no perturbation variables in the
conventional sense. While the Ellis-Bruni quantities are adapted to  
comoving
hypersurfaces, the new variables are associated to constant expansion
hypersurfaces, i.e., hypersurfaces of uniform Hubble parameter,
hypersurfaces of constant 3-curvature, and constant density  
hypersurfaces.

We shall establish the connections of these new quantities both to the
Ellis-Bruni variables and among themselves. We will reconsider the
perturbation theory of a perfect fluid universe and we will show  
that the
corresponding analysis formally simplifies if written in terms of  
the new
variables. This simplification is especialy useful with respect to the
characterization of a conserved quantity in the limit of large  
perturbation
scales.

For a flat background we find expressions for the latter in terms of the
fractional spatial energy density gradient (or the spatial gradient  
of the
3-curvature) on constant expansion hypersurfaces, or, alternatively, in
terms of the expansion (or 3-curvature) gradients on constant density
hypersurfaces. In the case of nonvanishing background curvature none of
these quantities will be conserved, in general. Only for dust or  
under the
additional condition that the curvature terms in the background field
equations are small, the above quantities remain approximately  
constant. The
most suitable quantity to characterize this case turns out to be the
fractional, spatial gradient of the energy density on hypersurfaces of
constant 3-curvature.

The paper is organized as follows: In section II we recall the  
gauge-problem
in conventional perturbation theory and point out the relation  
between the
gauge-invariant perturbation variables used previously by the  
author \cite
{ZCQG,ZApJ} and the Ellis-Bruni variables \cite{EB}. In section III we
consider the cosmological dynamics in terms of the Ellis-Bruni  
variables as
modified by Jackson \cite{Jack}. In section IV we define covariant and
gauge-invariant variables associated with constant expansion  
hypersurfaces,
i.e. hypersurfaces of constant Hubble parameter, hypersurfaces of  
constant
curvature and constant density hypersurfaces, and we establish relations
between them. In section V different expressions for large-scale  
conserved
quantities for vanishing background curvature are found and the  
problems to
generalize these results to curved backgrounds are discussed. Section VI
presents the conclusions of the paper.

\section{The gauge problem}

The investigation of cosmological perturbations within general  
relativity
was pioneered by Lifshitz \cite{Lif}. The traditional approach  
(see, e.g. 
\cite{Weinberg}) starts with Einstein's field equations 
\begin{equation}
G _{mn} \equiv R _{mn} - \frac{1}{2}g _{mn} R = \kappa T _{mn}   
\label{1}
\end{equation}
$\left(m, n .... = 0,1,2,3 \right)$ and splits both sides of the  
latter into
a zeroth-order background and first-order perturbations about this
background: 
\begin{equation}
G _{mn} = G _{mn}^{\left(0\right)} + \hat{G}_{mn}\ ,\ \ \ T _{mn} = T
_{mn}^{\left(0\right)} + \hat{T}_{mn}\ .\ \ \   \label{2}
\end{equation}
Provided, the zeroth-order problem $G _{mn}^{\left(0\right)} = \kappa T
_{mn}^{\left(0\right)} $ is solved, one has to study the set of  
perturbation
equations 
\begin{equation}
\hat{G}_{mn} = \kappa \hat{T}_{mn}\ .  \label{3}
\end{equation}
Restricting ourselves to a perfect fluid with the energy-momentum  
tensor 
\begin{equation}
T _{mn} = \rho u _{m}u _{n} + p h _{mn}\ ,  \label{4}
\end{equation}
where $\rho $ is the energy density, $p$ is the pressure, $u ^{m}$  
is the
4-velocity and $h^{mn} = g^{mn} + u^{m}u^{n}$, the equations (\ref{3})
represent a system of differential equations for the perturbation  
quantities 
\begin{equation}
\hat{\rho } \equiv \rho - \rho ^{\left(0\right)}\ ,\ \ \hat{p}  
\equiv p - p
^{\left(0\right)}\ ,\ \ \hat{u}_{m} \equiv u _{m}- u  
_{m}^{\left(0\right)}\
,\ \ \hat{g}_{mn} \equiv g _{mn} - g _{mn}^{\left(0\right)}\ .\ \    
\label{5}
\end{equation}
None of these quantities is gauge-invariant, i.e., invariant under
infinitesimal coordinate transformations 
\begin{equation}
x ^{n \prime} = x ^{n} - \zeta ^{n}\left(x\right)\ .  \label{6}
\end{equation}
Assuming a homogeneous and isotropic, comoving (i.e., $u ^{m  
\left(0\right)}
= (1,0,0,0)$,) zeroth order, the transformation properties of these
quantities are (see, e.g., \cite{Weinberg}, chapter 10.9) 
\begin{equation}
\hat{\rho }^{\prime } = \hat{\rho } + \dot{\rho }^{\left(0 \right)}\zeta
^{0}\ ,\ \ \ \hat{p}^{\prime } = \hat{p} + \dot{p}^{\left(0  
\right)}\zeta
^{0}\ ,\ \ \ \hat{u}_{m}^{\prime } = \hat{u}_{m} - \zeta ^{0}_{,m}\ ,\ 
\label{7}
\end{equation}
where we have used $\rho _{,0}^{\left(0 \right)} = \dot{\rho }^{\left(0
\right)} \equiv \rho _{,n}^{\left(0 \right)} u ^{n \left(0  
\right)}$ etc. in
the comoving zeroth order, and 
\begin{equation}
\hat{g} _{mn}^{\prime } = \hat{g}_{mn} + \zeta ^{a}_{,m}g _{an}^{\left(0
\right)} + \zeta ^{a}_{,n}g _{ma}^{\left(0 \right)} + \zeta ^{a}g
_{mn,a}^{\left(0 \right)}\ .  \label{8}
\end{equation}

One may either chose now a specific gauge, e.g., the traditionally  
preferred
synchronous gauge, or try to find a gauge-invariant desciption (\cite
{Bardeen}). In the latter case one may look for suitable  
combinations of the
quantities (\ref{5}) that are gauge-invariant, i.e., invariant under the
infinitesimal coordinate transformations (\ref{6}). From the  
transformation
properties (\ref{7}) it is obvious that, e.g., the quantity 
\begin{equation}
\hat{\rho}^{\left(c\right)}_{,\mu } \equiv \hat{\rho}_{,\mu } +  
\dot{\rho }
^{\left(0 \right)}\hat{u}_{\mu }  \label{9}
\end{equation}
$\left(\mu ... = 1,2,3 \right)$ is invariant under the  
transformations (\ref
{6}), i.e. $\hat{\rho}^{\left(c\right)\prime }_{,\mu } = \hat{\rho}
^{\left(c\right)}_{,\mu }$. This quantity represents the energy density
perturbations on comoving (superscript `$c$') hypersurfaces  
$\hat{u} _{\mu }
= 0$ (cf \cite{ZCQG,ZApJ}). It corresponds to the spatial derivative of
Bardeen's quantity $\epsilon _{m}$ \cite{Bardeen,BDE}.

One easily recognizes that this construction principle of  
gauge-invariant
quantities is applicable for any scalar (\cite{ZCQG,ZApJ}). E.g.,  
pressure
perturbations on comoving hypersurfaces are characterized by 
\begin{equation}
\hat{p}^{\left(c\right)}_{,\mu } \equiv \hat{p}_{,\mu } +  
\dot{p}^{\left(0
\right)}\hat{u}_{\mu }\ .  \label{10}
\end{equation}
Gauge-invariant descriptions of scalars such as the fluid expansion  
$\Theta
\equiv u ^{a}_{;a}$ and the spatial 3-curvature ${\cal R}$, to be  
defined
below, are 
\begin{equation}
\hat{\Theta }^{\left(c\right)}_{,\mu } = \hat{\Theta }_{,\mu } +  
\dot{\Theta 
}^{\left(0 \right)}\hat{u}_{\mu }\ ,\ \ \ \hat{{\cal  
R}}^{\left(c\right)}_{,
\mu } = \hat{{\cal R}}_{,\mu } + \dot{{\cal R}}^{\left(0 \right)}\hat{u}
_{\mu }\ ,  \label{11}
\end{equation}
respectively. All these quantities are gauge-invariant by  
construction. In
order to establish a link between these variables and those  
introduced by
Ellis and Bruni \cite{EB}, we have to investigate the key  
quantities of the
latter authors, the spatially projected gradients $h ^{c}_{a}\rho  
_{,c}$, $h
^{c}_{a} p_{,c}$, $h ^{c}_{a}\Theta _{,c}$, $h ^{c}_{a} {\cal R}_{,c}$,
etc., in first order. Using the first-order expressions 
\begin{equation}
\hat{h}^{\alpha}_{\beta } = 0 \ ,\ \ \ \ \hat{h}^{0}_{\alpha} = u  
_{\alpha}\
,  \label{12}
\end{equation}
one finds 
\begin{equation}
\left(h ^{c}_{0}\rho _{,c}\right)^{\hat{}} = 0 \ ,\ \ \ \ \ \left(h
^{c}_{\mu }\rho _{,c}\right)^{\hat{}} = \hat{\rho}_{,\mu } + \dot{\rho }
^{\left(0 \right)}\hat{u} _{\mu } \equiv \hat{\rho  
}^{\left(c\right)}_{,\mu
}\ ,  \label{13}
\end{equation}
and corresponding relations for the spatial gradients of $p$,  
$\Theta $, and 
${\cal R}$ in first order. While the quantities $h ^{c}_{a}\rho  
_{,c}$ etc.
vanish in zeroth order since the background was assumed to be  
homogeneous,
they coincide in first order with the gauge-invariant quantities  
$\hat{\rho}
^{\left(c\right)}_{,\mu }$ etc.

The second relation (\ref{13}) demonstrates explicitly how the  
Ellis-Bruni
variables are related to corresponding gauge-invariant perturbation
quantities with physical interpretations on comoving hypersurfaces. Our
considerations may be regarded as a kind of motivation for the  
usefulness of
the Ellis-Bruni quantities from the point of view of conventional
perturbation theory. They also establish the connection beween the  
latter
quantities and the quantities of the type (\ref{9}) and (\ref{10}),
introduced in \cite{ZCQG,ZApJ}. Using the Ellis-Bruni quantities  
which, by
their definition, vanish in a homogeneous universe, one may avoid an
explicit decomposition into a fictitious background and  
perturbations about
this (not unique) zeroth order. Since the existence of a  
gauge-problem is
just a consequence of the nonuniqueness of the background, one  
circumvents
the gauge-problem from the beginning by using the Ellis-Bruni variables
which are covariant by definition and no perturbation quantities in the
usual sense.

The spatially projected gradients by Ellis and Bruni have a  
physical meaning
on comoving hypersurfaces. As to the physical understanding of the
gauge-invariant quantities $h ^{c}_{a}\rho _{,c}$ etc., if regarded as
conventional first-order perturbation variables in the sense described
above, the comoving gauge is naturally preferred. In different  
gauges, these
quantities have no obvious physical meaning, in general. From these
statements the impression may arise that there is something special  
about
the comoving gauge, even within a gauge-invariant formalism. This  
is not the
case, however, as we are going to clarify in this paper. It is  
well-known
that there exist obviously reasonable gauge-invariant quantities in the
literature having a simple physical meaning in gauges different from the
comoving gauge. For reasons of convenience some authors e.g., Bardeen,
Steinhardt and Turner \cite{BST}, occasionally prefer the `uniform  
Hubble
constant gauge' or others, e.g., Hwang \cite{Hwang}, the constant  
curvature
gauge. One may therefore ask whether there exist covariant and
gauge-invariant quantities that, from the point of view of their  
physical
interpretation, prefer hypersurfaces of constant Hubble parameter or of
constant curvature in a similar sense in which the Ellis-Bruni  
quantities
prefer comoving hypersurfaces. As we will see below, there are indeed
problems for which a choice of covariant variables different from the
Ellis-Bruni choice may simplify the dynamical description. In order to
introduce these variables we first recall the basic elements of the
cosmological dynamics in terms of the Ellis-Bruni quantities,  
following here
the elegant presentation by Jackson \cite{Jack}.

\section{Cosmological dynamics in terms of the Ellis-Bruni-Jackson  
variables}

Instead of applying the field equations (\ref{1}), we shall  
investigate the
cosmological dynamics within the `fluid-flow' approach used in \cite
{Hawk,Ols,Lyth,LyMu,LyStew}. The equations of motion $T ^{ik}_{\ ;  
k} = 0$,
imply 
\begin{equation}
\dot{\rho } = - \Theta\left(\rho + p\right) \ ,  \label{14}
\end{equation}
with 
\begin{equation}
\Theta \equiv u^{i}_{\ ;i}\ ,  \label{15}
\end{equation}
and 
\begin{equation}
\left(\rho + p\right)\dot{u}_{}^{m} = - p_{,k}h^{mk}\ ,  \label{16}
\end{equation}
where $\dot{u}^{m} \equiv u^{m}_{;n}u^{n}$. Additionally, we shall  
use the
Raychaudhuri equation for $\Theta $, 
\begin{equation}
\dot{\Theta} + \frac{1}{3}\Theta^{2} + 2\left(\sigma^{2} -  
\omega^{2}\right)
- \dot{u}^{a}_{;a} - \Lambda + \frac{\kappa}{2}\left(\rho +  
3p\right) = 0 \ .
\label{17}
\end{equation}
$\Lambda $ is the cosmological constant. The magnitudes of shear and
vorticity are defined by 
\begin{equation}
\sigma^{2} \equiv \frac{1}{2}\sigma_{ab}\sigma^{ab}\ ,\ \ \ \ \  
\omega^{2}
\equiv \frac{1}{2}\omega_{ab}\omega^{ab}\ ,  \label{18}
\end{equation}
with 
\begin{equation}
\sigma_{ab} = h_{a}^{c}h_{b}^{d}u_{\left(c;d\right)} - \frac{1}{3}
\Theta_{}h_{ ab}\ ,\ \ \ \ \ \omega_{ab} =
h_{a}^{c}h_{b}^{d}u_{\left[c;d\right]} \ .  \label{19}
\end{equation}
The 3-curvature scalar of the projected metric, 
\begin{equation}
{\cal R} = 2 \left(- \frac{1}{3}\Theta ^{2} + \sigma ^{2} - \omega  
^{2} +
\kappa \rho + \Lambda \right)\ ,  \label{20}
\end{equation}
reduces to the 3-curvature of the surfaces orhogonal to $u ^{a}$ in  
the case 
$\omega = 0$. The homogeneous and isotropic FLRW universes are  
characterized
by $\sigma = \omega = \dot{u}^{a} = 0$. Taking the spatial gradient  
of (\ref
{14}) yields 
\begin{equation}
h ^{c}_{a}\dot{\rho }_{,c} = - \left(\rho + p\right)h  
^{c}_{a}\Theta _{,c} -
\Theta h ^{c}_{a}\left(\rho + p\right)_{,c}\ .  \label{21}
\end{equation}
Introducing a length scale $S$ by 
\begin{equation}
\frac{1}{3}\Theta \equiv \frac{\dot{S}}{S}  \label{22}
\end{equation}
and rewriting the l.h.s. of (\ref{21}) after multiplying by $S$ as 
\begin{equation}
Sh ^{c}_{m}\dot{\rho }_{,c} = h ^{a}_{m}\left(Sh ^{c}_{a}\rho _{,c}  
\right)^{
\displaystyle \cdot} - \Theta S h ^{c}_{m}p _{,c} + \left(\omega  
^{c}_{m} +
\sigma ^{c}_{m}\right)Sh ^{n}_{c} \rho _{,n} \ ,  \label{23}
\end{equation}
the equations (\ref{21}) and (\ref{23}) may be combined into 
\begin{equation}
h _{n}^{a} \dot{D}_{a} + \frac{\dot{p}}{\rho + p} D _{n} + \left(\omega
^{c}_{n} + \sigma ^{c}_{n}\right) D _{c} + t _{n} = 0 \ ,  \label{24}
\end{equation}
where we have introduced the fractional quantity \cite{Jack} 
\begin{equation}
D _{a} \equiv \frac{S h ^{c}_{a}\rho _{,c}}{\rho + p} \ ,  \label{25}
\end{equation}
and 
\begin{equation}
t _{a} \equiv S h ^{c}_{a} \Theta _{,c} \ .  \label{26}
\end{equation}
Equation (\ref{24}) is Jackson's eq.(29) \cite{Jack}. Similarly,  
taking the
spatially projected gradient of the Raychaudhuri equation  
(\ref{17}), one
obtains 
\begin{eqnarray}
h _{n}^{a} \dot{t}_{a}&=& - \dot{\Theta }P _{n} - \left(\omega  
^{c}_{n} +
\sigma ^{c}_{n}\right) t _{c} - \frac{2}{3} \Theta t _{n} - S h ^{c}_{n}
\left(2 \sigma ^{2} - 2 \omega ^{2}\right)_{,c}  \nonumber \\
&& + S h ^{c}_{n} \left( \dot{u}^{a}_{;a}\right)_{,c} -  
\frac{\kappa }{2}
\left(\rho + p\right) \left[D _{n} + 3 P _{n}\right] \ ,  \label{27}
\end{eqnarray}
with the fractional quantity 
\begin{equation}
P _{a} \equiv \frac{S h ^{c}_{a}p _{,c}}{\rho + p} \ ,  \label{28}
\end{equation}
characterizing the pressure perturbations. Equation (\ref{27}) is  
Jackson's
eq.(30) \cite{Jack}. The set of equations (\ref{24}) and (\ref{27})  
is still
completely general. Even for an equation of state $p = p \left(\rho  
\right)$
which allows one to express $P _{n}$ in terms of $D _{n}$, the  
equations (
\ref{24}) and (\ref{27}) are, of course, not a closed system for $D  
_{a}$
and $t _{a}$ since these quantities are coupled to $\omega $ and  
$\sigma $
and their spatial gradients.

From now on we shall assume the spatial gradients as well as  
$\sigma $ and $
\omega $ to be small, i.e., we assume the universe to be almost  
homogeneous
and isotropic. Consequently, up to first order in the  
inhomogeneities, the
factors in front of the quantities $D _{a}$, $P _{a}$ and $t _{a}$  
in (\ref
{24}) and (\ref{27}) refer to the homogeneous and isotropic case with $
\omega = \sigma = 0$. The linearized set of equations becomes 
\begin{equation}
h _{n}^{a} \dot{D}_{a} + \frac{\dot{p}}{\rho + p} D _{n} + t _{n} =  
0 \ ,
\label{29}
\end{equation}
and 
\begin{equation}
h _{n}^{a} \dot{t}_{a} = - \frac{1}{2} {\cal R} P _{n} -  
\frac{2}{3} \Theta
t _{n} - \frac{\kappa }{2} \left(\rho + p\right)D _{n} + S h  
^{c}_{n} \left( 
\dot{u}^{a}_{;a}\right)_{,c} \ ,  \label{30}
\end{equation}
where we have used the zeroth-order relations ( $\Lambda = 0$) 
\begin{equation}
\kappa \rho = \frac{1}{3}\Theta ^{2} + \frac{1}{2}{\cal R} \ ,   
\label{31}
\end{equation}
and 
\begin{equation}
\dot{\Theta } + \frac{3}{2}\kappa \left(\rho + p\right) =  
\frac{1}{2} {\cal R
} \ ,  \label{32}
\end{equation}
for homogeneous and isotropic universes. The 3-curvature in the  
latter case
is known to be 
\begin{equation}
{\cal R} = \frac{6 k}{a ^{2}} \ ,  \label{33}
\end{equation}
with the scale factor $a$ of the Robertson-Walker metric. The last  
term in (
\ref{30}) is generally given by (\cite{Jack}) 
\begin{equation}
S h ^{c}_{n} \left( \dot{u}^{a}_{;a}\right)_{,c} = - S h ^{c}_{n}\left[h
^{ab} \left(h _{a}^{m}\frac{p _{,m}}{\rho + p}\right)_{;b}  
\right]_{,c} + S
h ^{c}_{n}\left[h ^{ab} \frac{p _{,a}}{\rho + p}\frac{p _{,b}}{\rho + p}
\right]_{,c} \ .  \label{34}
\end{equation}
In linear order in the inhomogeneities this reduces to 
\begin{equation}
S h ^{c}_{n} \left( \dot{u}^{a}_{;a}\right)_{,c} = - \frac{\nabla  
^{2}}{a
^{2}} P _{n}\ .  \label{35}
\end{equation}
With an equation of state $p = p \left(\rho \right)$ one may write 
\begin{equation}
\dot{p} = - c _{s}^{2}\Theta \left( \rho + p\right)\ , \ \ \ \ \ P  
_{n} = c
_{s}^{2}D _{n} \ ,  \label{36}
\end{equation}
where $c _{s}^{2} \equiv \left(\partial p/\partial \rho  
\right)_{ad}$ is the
square of the sound velocity. Because of (\ref{13}) etc., in linear  
order,
the system (\ref{29}), (\ref{30}) may be written as 
\begin{equation}
\dot{D}_{\mu } - c _{s}^{2} \Theta D _{\mu } + t _{\mu } = 0 \ ,   
\label{37}
\end{equation}
and 
\begin{equation}
\dot{t}_{\mu } = - \frac{2}{3} \Theta t _{\mu } -  
\left[\frac{\kappa }{2}
\left(\rho + p\right) + 3 \frac{k}{a ^{2}} c _{s}^{2} + c _{s}^{2}\frac{
\nabla ^{2}}{a ^{2}} \right] D _{\mu } \ .  \label{38}
\end{equation}
Eliminating $t _{\mu }$ from the system (\ref{37}), (\ref{38}), one  
obtains
a closed equation for $D _{\mu }$, 
\begin{eqnarray}
\ddot{D}_{\mu } &+& \left(\frac{2}{3} - c _{s}^{2}\right) \Theta \dot{D}
_{\mu }  \nonumber \\
&& - \left[\left(c_{s}^{2} \right)^{\displaystyle \cdot}\Theta + \frac{
\kappa }{2}\left(\rho - 3 p\right) c _{s}^{2} + \frac{\kappa }{2}  
\left(\rho
+ p\right) + c _{s}^{2} \frac{\nabla ^{2}}{a ^{2}}\right] D _{\mu }  
= 0 \ ,
\label{39}
\end{eqnarray}
which corresponds to Jackson's eqation (57). From the latter  
equation one
easily derives the well-known growing and decaying modes in the
long-wavelength limit for $k = 0$, equivalent to neglecting the spatial
gradient term in the bracket in front of $D _{\mu }$.

\section{New perturbation variables}

Let us consider the ratio of the spatial variation of a scalar quantity,
say, the energy density, i.e., $h ^{c}_{a}\rho _{,c}$ to its  
variation in
time, $\dot{\rho } \equiv u ^{c}\rho _{,c}$. With (\ref{13}) we find, in
linear order, 
\begin{equation}
\left(\frac{h ^{c}_{\mu }\rho _{,c}}{ u ^{c}\rho  
_{,c}}\right)^{\hat{}} = 
\frac{\hat{\rho }_{,\mu }}{\dot{\rho }} + \hat{u}_{\mu }\ .  \label{40}
\end{equation}
Obviously, this ratio is a reasonable quantity to characterize small
deviations from homogeneity. Similar ratios may be formed for the other
scalar quantities of interest: 
\begin{equation}
\left(\frac{h ^{c}_{\mu }p _{,c}}{ u ^{c}p_{,c}}\right)^{\hat{}} = \frac{
\hat{p}_{,\mu }}{\dot{p }} + \hat{u}_{\mu }\ ,  \label{41}
\end{equation}
and 
\begin{equation}
\left(\frac{h ^{c}_{\mu }\Theta _{,c}}{ u ^{c}\Theta  
_{,c}}\right)^{\hat{}}
= \frac{\hat{\Theta }_{,\mu }}{\dot{\Theta }} + \hat{u}_{\mu }\ .   
\label{42}
\end{equation}
For $k \neq 0$ we may also introduce 
\begin{equation}
\left(\frac{h ^{c}_{\mu }{\cal R} _{,c}} { u ^{c} {\cal R}  
_{,c}}\right) ^{
\hat{}} = \frac{{\hat{{\cal R}}}_{,\mu }}{\dot{{\cal R} }} +  
\hat{u}_{\mu }\
\ \ \ \ \left(k \neq 0 \right)\ .  \label{43}
\end{equation}
For $k = {\cal R} = 0$ the 3-curvature perturbations are  
gauge-invariant,
i.e., 
\begin{equation}
\left(h ^{c}_{\mu } {\cal R} _{,c} \right) ^{\hat{}} = \hat{{\cal  
R}}_{,\mu
}\ \ \ \ \ \left(k = 0 \right)\ .  \label{44}
\end{equation}

By construction, all the quantities (\ref{40}) - (\ref{43})  
represent the
first-order ratio of the spatial variation of the quantities $\rho  
$, $p$, $
\Theta $ and ${\cal R}$, respectively, to their change in time, on  
comoving
hypersurfaces $\hat{u}_{\mu } = 0$. The 4-velocity perturbation $\hat{u}
_{\mu }$ enters each of the relations (\ref{40}) - (\ref{43}) in  
exactly the
same manner, namely simply additively. Forming differences between  
any two
of the quantities (\ref{40}) - (\ref{43}), the velocity  
perturbations just
cancel. This suggests combining (\ref{40}) - (\ref{43}) in order to  
obtain
new, covariant and gauge-invariant variables. E.g., we may define 
\begin{equation}
\frac{h ^{c}_{m }\rho _{,c}^{\left(ce \right)}} { u ^{c}\rho _{,c}}  
\equiv 
\frac{h ^{c}_{m }\rho _{,c}}{ u ^{c}\rho _{,c}} - \frac{h ^{c}_{m  
}\Theta
_{,c}}{ u ^{c}\Theta _{,c}} \ , \ \ \ \ \ \ \left(k = 0, \pm 1\right)\ ,
\label{45}
\end{equation}
or 
\begin{equation}
\frac{h ^{c}_{m}\rho _{,c}^{\left(cc \right)}} { u ^{c}\rho _{,c}}  
\equiv 
\frac{h ^{c}_{m }\rho _{,c}}{ u ^{c}\rho _{,c}} - \frac{h ^{c}_{m}  
{\cal R}
_{,c}} { u ^{c} {\cal R} _{,c}} \ \ \ \ \ \left(k = \pm 1\right)\ .
\label{46}
\end{equation}
In linear order, with (\ref{40}), (\ref{42}) and (\ref{43}), the spatial
components are 
\begin{equation}
\left(\frac{h ^{c}_{\mu }\rho _{,c}^{\left(ce \right)}} { u  
^{c}\rho _{,c}}
\right)^{\hat{}} = \frac{\hat{\rho }_{,\mu }}{\dot{\rho }} - \frac{\hat{
\Theta }_{,\mu }}{\dot{\Theta }} \ \ \ \ \left(k = 0, \pm 1\right) \ ,
\label{47}
\end{equation}
and 
\begin{equation}
\left(\frac{h ^{c}_{\mu }\rho _{,c}^{\left(cc \right)}} { u  
^{c}\rho _{,c}}
\right)^{\hat{}} = \frac{\hat{\rho }_{,\mu }}{\dot{\rho }} - \frac{\hat{
{\cal R}}_{,\mu }}{\dot{{\cal R} }} \ \ \ \ \left(k = \pm 1\right) \ .
\label{48}
\end{equation}
Generally, all variables of the type (\ref{45}) and (\ref{46}) are exact
covariant quantities without any reference to perturbation theory. As
perturbation variables they are gauge-invariant by construction.  
From (\ref
{47}) it seems obvious to conclude that $h ^{c}_{m}\rho _{,c}^{\left(ce
\right)} / \left(u ^{c}\rho _{,c} \right)$ in first order represents the
ratio of spatial to time variation of the energy density on  
hypersurfaces $
\hat{\Theta }_{,\mu } = 0$, i.e., on hypersurfaces of constant  
expansion or
constant Hubble parameter, in the same sense as the Ellis-Bruni  
quantity $h
^{c}_{m}\rho _{,c} / \left(u ^{c}\rho _{,c} \right)$ has a  
corresponding,
well-known meaning in the comoving gauge $\hat{u}_{\mu } = 0$  
(cf.(\ref{13}
)). While we shall indeed use this interpretation of (\ref{47}) in the
following, it requires additional clarification. The point is that the
quantity $\rho $, introduced by (\ref{4}), is the energy density for a
comoving (with 4-velocity $u ^{a}$) observer. A different observer,  
moving,
e.g., with a 4-velocity $n ^{a}$, normal to hypersurfaces  
$\hat{\Theta }_{,
\mu } = 0$, would interpret a different quantity, namely $\mu = T _{ab}n
^{a}n ^{b}$ as energy density. (Different from the observer moving  
with $u
^{a}$ he would also measure an energy flux.) Generally, $\rho $ and  
$\mu $
are related by (see \cite{KiEll,BDE}) 
\[
\mu = \rho \cosh ^{2}\beta + p \sinh ^{2}\beta \ , 
\]
where $\beta (t)$ is the hyperbolic angle of tilt given by $\cosh  
\beta = -
u ^{a}n _{a}$. Corresponding relations hold for the other  
quantities like
pressure and expansion. It follows that the perturbation quantity  
$\hat{\rho 
}$ for a comoving (with $u ^{a}$) observer will generally not  
coincide with
the perturbation $\hat{\mu } $, i.e., $\hat{\rho }$ will not  
coincide with
the energy density perturbation on hypersurfaces of constant  
expansion. Now,
in the present case we are considering first-order perturbations,  
while $u
^{a}$ and $n ^{a}$ coincide in zeroth order. For small angles of tilt,
however, i.e., for $\beta \ll 1$, the differences between $\mu $  
and $\rho $
are of second order in $\beta $. Consequently, within linear  
perturbation
theory we may identify the quantities $\hat{\mu }$ and $\hat{\rho }$.
Similar statements hold for the other perturbation variables.  
Therefore, the
above interpretation of regarding the quantity (\ref{47}) as the  
ratio of
spatial to time change of the energy density on hypersurfaces of  
constant
expansion, $\hat{\Theta }_{,\mu } = 0$, is justified up to first order.
Since the latter gauge is defined by the vanishing of the gradient  
of $\hat{
\Theta }$, the quantity $\hat{\Theta }$ itself is only determined  
up to a
constant.

Analogously, for $k = \pm 1$, the quantity $h ^{c}_{m}\rho  
_{,c}^{\left(cc
\right)} / \left(u ^{c}\rho _{,c} \right)$ in first order represents the
ratio of spatial to time change of the energy density on  
hypersurfaces of
constant curvature (superscript `cc'), defined by $\hat{{\cal  
R}}_{,\mu } = 0
$. The same construction principle may be applied to introduce a  
variable
describing the ratio of spatial to time change of the 3-curvature, 
\begin{equation}
\frac{h ^{c}_{m}{\cal R} ^{\left(ce\right)} _{,c}} { u ^{c}{\cal R}  
_{,c}}
\equiv \frac{h ^{c}_{m}{\cal R} _{,c}} { u ^{c}{\cal R} _{,c}} - \frac{h
^{c}_{m}\Theta _{,c}} { u ^{c}\Theta _{,c}} \ \ \ \ \left(k = \pm  
1\right) \
.  \label{49}
\end{equation}
For $k = 0$ one has $h ^{c}_{\mu }{\cal R}_{,c}^{\left(ce\right)} = h
^{c}_{\mu }{\cal R}_{,c} $.

The ratio of spatial to time change of the expansion with respect to the
corresponding ratio for the 3-curvature is 
\begin{equation}
\frac{h ^{c}_{m }\Theta _{,c}^{\left(cc\right)}} { u ^{c}\Theta  
_{,c}} = 
\frac{h ^{c}_{m }\Theta _{,c}} { u ^{c}\Theta _{,c}} - \frac{h  
^{c}_{m}{\cal 
R} _{,c}} { u ^{c}{\cal R} _{,c}} \ \ \ \ \left(k = \pm 1\right)\ .
\label{50}
\end{equation}
Again, the 0-component vanishes and the first-order spatial  
components may
be written analogously to (\ref{47}) and (\ref{48}), namely 
\begin{equation}
\left(\frac{h ^{c}_{\mu } {\cal R} _{,c}^{\left(ce \right)}} { u  
^{c}{\cal R}
_{,c}}\right)^{\hat{}} = \frac{\hat{{\cal R} }_{,\mu }}{\dot{{\cal  
R}}} - 
\frac{\hat{\Theta }_{,\mu }}{\dot{\Theta }} \ \ \ \ \left(k = \pm  
1\right)\ ,
\label{51}
\end{equation}
and 
\begin{equation}
\left(\frac{h ^{c}_{\mu }\Theta _{,c}^{\left(cc \right)}} { u ^{c}\Theta
_{,c}}\right)^{\hat{}} = \frac{\hat{\Theta }_{,\mu }}{\dot{\Theta  
}} - \frac{
\hat{{\cal R}}_{,\mu }}{\dot{{\cal R}}} \ \ \ \ \left(k = \pm  
1\right) \ .
\label{52}
\end{equation}
The quantity (\ref{51}) represents the first-order ratio of spatial  
to time
change of the 3-curvature on constant expansion hypersurfaces,  
while (\ref
{52}) is the corresponding ratio for the expansion on constant curvature
hypersurfaces.

Provided, an equation of state $p = p \left(\rho \right)$ is given,  
we are
left with three quantities being ratios of spatial to time changes,  
namely
those for $\rho $, $\Theta $ and ${\cal R}$. Forming differences between
them yields variables that in linear order may be interpreted as
perturbation variables, e.g., on constant expansion hypersurfaces  
like (\ref
{47}) and (\ref{51}), or, like (\ref{48}) and (\ref{52}), on  
hypersurfaces
of constant curvature. In a similar way it is possible to introduce
variables which in first order have physical interpretations on constant
density hypersurfaces. The corresponding definitions are 
\begin{equation}
\frac{h ^{c}_{m} {\cal R}^{\left(cd\right)} _{,c}} { u ^{c}{\cal R}  
_{,c}}
\equiv \frac{h ^{c}_{m}{\cal R} _{,c}} { u ^{c}{\cal R} _{,c}} - \frac{h
^{c}_{m}\rho _{,c}} { u ^{c}\rho _{,c}} \ \ \ \ \ \ \ \ \left(k = \pm
1\right)\ ,  \label{53}
\end{equation}
and 
\begin{equation}
\frac{h ^{c}_{m }\Theta _{,c}^{\left(cd\right)}} { u ^{c}\Theta  
_{,c}} = 
\frac{h ^{c}_{m }\Theta _{,c}} { u ^{c}\Theta _{,c}} - \frac{h  
^{c}_{m}\rho
_{,c}} { u ^{c}\rho _{,c}} \ \ \ \ \ \left(k = 0, \pm 1\right)\ ,   
\label{54}
\end{equation}
where the superscript `cd' stands for `constant density'. In linear  
order
one has 
\begin{equation}
\left(\frac{h ^{c}_{\mu }{\cal R} _{,c}^{\left(cd \right)}} { u  
^{c}{\cal R}
_{,c}}\right)^{\hat{}} = \frac{\hat{{\cal R} }_{,\mu }}{\dot{{\cal  
R}}} - 
\frac{\hat{\rho }_{,\mu }}{\dot{\rho }} \ \ \ \ \left(k = \pm  
1\right)\ ,
\label{55}
\end{equation}
and 
\begin{equation}
\left(\frac{h ^{c}_{\mu }\Theta _{,c}^{\left(cd \right)}} { u ^{c}\Theta
_{,c}}\right)^{\hat{}} = \frac{\hat{\Theta }_{,\mu }}{\dot{\Theta  
}} - \frac{
\hat{\rho }_{,\mu }}{\dot{\rho }} \ \ \ \ \left(k = 0, \pm 1\right)\ .
\label{56}
\end{equation}
The quantity (\ref{55}) is (\ref{48}) with the opposite sign. A
corresponding relation holds between (\ref{56}) and (\ref{47}).

The relations of all of these quantities to the Ellis-Bruni  
variables are
obvious. It will turn out that the perturbation dynamics looks  
simpler in
any of the new variables than it does in terms of the Ellis-Bruni  
variables.
The reason is that the new variables are adapted to hypersurfaces  
on which
one of the first-order spatial gradients of $\Theta $, ${\cal R}$  
or $\rho $
vanishes. None of these gradients vanishes, however, on comoving
hypersurfaces.

\subsection{Constant expansion hypersurfaces}

Differentiating the Gau\ss-Codazzi equation (\ref{20}) and projecting
orthogonal to $u ^{a}$ yields, in linear order in the inhomogeneities, 
\begin{equation}
r _{a} = - \frac{4}{3}\Theta t _{a} + 2 \kappa \left(\rho +  
p\right) D _{a}
\ ,  \label{57}
\end{equation}
where we have introduced the abbreviation 
\begin{equation}
r _{a} \equiv S h _{a}^{c}{\cal R} _{,c}\ .  \label{58}
\end{equation}
According to (\ref{45}) and (\ref{47}) we introduce the fractional  
quantity 
\begin{equation}
D _{a}^{\left(ce\right)} \equiv D _{a} - \frac{\dot{\rho }}{\rho +  
p} \frac{
t _{a}}{\dot{\Theta }} \ \ \ \ \ \ \ \ \left(k = 0, \pm 1\right)\ \ \ ,
\label{59}
\end{equation}
which in linear order represents the fractional, spatial gradient of the
energy density on hypersurfaces of constant expansion. Analogously, we
define the fractional spatial gradient of the curvature on constant
expansion hypersurfaces, 
\begin{equation}
r _{a}^{\left(ce\right)} \equiv r _{a} - \dot{{\cal R}}\frac{t  
_{a}}{\dot{
\Theta }}\ \ \ \ \ \ \ \left(k = \pm 1\right)\ ,  \label{60}
\end{equation}
with (cf (\ref{33})) 
\begin{equation}
\dot{{\cal R}} = - \frac{2}{3}\Theta\ {\cal R}\ .  \label{61}
\end{equation}
Using the definitions (\ref{59}) and (\ref{60}) to replace $D _{a}$  
and $t
_{a}$, respectively, in (\ref{57}), we find a relation between the  
gradients
of the 3-curvature and the fractional energy density on uniform Hubble
parameter hypersurfaces: 
\begin{equation}
r _{a}^{\left(ce\right)} = 2 \kappa \left(\rho + p\right) D  
_{a}^{\left(ce
\right)} \ ,\ \ \ \ \left(k = 0, \pm 1\right)\ .  \label{62}
\end{equation}
This is relation (\ref{57}) written in the variables that are adapted to
constant expansion hypersurfaces. The latter relation is also valid  
for $k =
0$, where $r _{a}^{\left(ce \right)} \stackrel{\left(k = 0  
\right)}{=} r _{a}
$.

\subsection{Constant curvature hypersurfaces}

According to (\ref{50}) and (\ref{52}) the fractional energy density
gradient on hypersurfaces of constant curvature is characterized by 
\begin{equation}
D _{a}^{\left(cc\right)} \equiv D _{a} - \frac{\dot{\rho }}{\rho +  
p} \frac{
r _{a}}{\dot{{\cal R}}}\ \ \ \ \left(k = \pm 1\right)\ .  \label{63}
\end{equation}
Analogously, for the gradient of the expansion on these  
hypersurfaces one
has 
\begin{equation}
t _{a}^{\left(cc \right)} \equiv t _{a} - \dot{\Theta  
}\frac{r_{a}}{\dot{
{\cal R}}}\ \ \ \ \left(k = \pm 1\right)\ .  \label{64}
\end{equation}
Inserting $D _{a}$ and $t _{a}$ from (\ref{63}) and (\ref{64}) into  
(\ref{57}
), and using the zeroth-order relations (\ref{32}) and (\ref{61}),  
we get 
\begin{equation}
\frac{4}{3}\Theta t _{a}^{\left(cc \right)} = 2 \kappa \left(\rho + p
\right) D _{a}^{\left(cc \right)}\ \ \ \ \left(k = \pm 1\right)\ ,
\label{65}
\end{equation}
which is again (\ref{57}), now in terms of variables adapted to constant
curvature hypersurfaces.

Equation (\ref{57}) is a linear relation between the three  
Ellis-Bruni type
perturbation quantities $D _{a}$, $t _{a}$ and $r _{a}$ defined  
with respect
to comoving hypersurfaces. The introduction of the variables  
(\ref{59}) and (
\ref{60}) or, (\ref{63}) and (\ref{64}), respectively, reduces  
(\ref{57}) to
relations (either (\ref{62}) or (\ref{65})) between only two variables,
adapted either to hypersurfaces of constant expansion or to those of
constant curvature. Effectively, the number of variables has been  
reduced in
both cases. One may choose, e.g., one of the fractional energy density
perturbations, $D _{a}^{\left(cc \right)}$ or $D  
_{a}^{\left(ce\right)}$ as
independent variable.

At any stage it is possible to change between the sets of variables
associated with either the comoving, or the uniform Hubble  
constant, or the
constant curvature hypersurfaces. Subtracting (\ref{63}) and (\ref{59})
yields 
\begin{equation}
D _{a}^{\left(cc \right)} - D _{a}^{\left(ce\right)} = -  
\frac{\dot{\rho }}{
\rho + p} \frac{r _{a}^{\left(ce \right)}}{\dot{{\cal R}}}\ \ \ \ \  
\ \ \
\left(k = \pm 1\right)\ .  \label{66}
\end{equation}
Using here the relation (\ref{62}) between $r _{a}^{\left(ce  
\right)}$ and $
D _{a}^{\left(ce \right)}$ as well as the zeroth-order relations  
(\ref{32})
and (\ref{61}), we find the following connection between the fractional
energy density gradients on constant curvature and constant expansion
hypersurfaces: 
\begin{equation}
\frac{1}{2}{\cal R} D _{a}^{\left(cc \right)} = \dot{\Theta } D
_{a}^{\left(ce \right)}\ \ \ \ \left(k = \pm 1\right)\ .  \label{67}
\end{equation}

\subsection{Constant density hypersurfaces}

We define, according to (\ref{54}) and (\ref{56}) with (\ref{26}), the
perturbations of the expansion on constant density hypersurfaces, 
\begin{equation}
t _{a}^{\left(cd \right)} \equiv t _{a} - \left(\rho + p  
\right)\dot{\Theta }
\frac{D _{a}}{\dot{\rho }}\ \ \ \ \ \left(k = 0, \pm 1\right)\ ,   
\label{68}
\end{equation}
as well as, according to (\ref{53}) and (\ref{55}) with (\ref{58}),
curvature perturbations on these hypersurfaces, 
\begin{equation}
r _{a}^{\left(cd \right)} \equiv r _{a} - \left(\rho + p  
\right)\dot{{\cal R}
}\frac{D _{a}}{\dot{\rho }}\ \ \ \ \ \left(k = \pm 1\right)\ .   
\label{69}
\end{equation}
Solving eq.(\ref{68}) for $t _{a}$, eq.(\ref{69}) for $r _{a}$ and  
inserting
into (\ref{57}) yields 
\begin{equation}
r _{a}^{\left(cd \right)} = - \frac{4}{3}\Theta t _{a}^{\left(cd  
\right)}\ \
\ \ \ \ \ \left(k = 0, \pm 1\right)  \label{70}
\end{equation}
for the perturbed Gau\ss-Codazzi equation in terms of variables that are
adapted to constant density hypersurfaces. The relation (\ref{70})  
is also
valid for $k = 0$, where $r _{a}^{\left(cd \right)}  
\stackrel{\left(k = 0
\right)}{=} r _{a}$. Introducing into (\ref{68}) the quantity $D
_{a}^{\left(ce \right)}$ by (\ref{59}), one obtains a relation  
between the
quantities $t _{a}^{\left(cd \right)}$ and $D _{a}^{\left(ce \right)}$, 
\begin{equation}
\Theta t _{a}^{\left(cd \right)} = \dot{\Theta } D _{a}^{\left(ce  
\right)} \
\ \ \ \left(k = 0, \pm 1\right)\ .  \label{71}
\end{equation}
Analogously, from (\ref{69}) and (\ref{63}) one gets 
\begin{equation}
r _{a}^{\left(cd \right)} = - \frac{2}{3}{\cal R} D _{a}^{\left(cc  
\right)}
\ \ \ \ \left(k = \pm 1\right)\ .  \label{72}
\end{equation}
It follows, that $t _{a}^{\left(cd \right)}$ and $D _{a}^{\left(cc  
\right)}$
are connected through 
\begin{equation}
\Theta t _{a}^{\left(cd \right)} = \frac{1}{2}{\cal R} D _{a}^{\left(cc
\right)}\ \ \ \ \left(k = \pm 1\right)\ .  \label{73}
\end{equation}

\section{Conserved quantities}

We are now going to demonstrate that the system (\ref{37}), (\ref{38})
becomes especially simple if rewritten in terms of the newly introduced
variables. Starting with the definition (\ref{63}) of the  
fractional energy
density gradient on hypersurfaces of constant curvature and using  
(\ref{14})
and (\ref{61}), we have 
\begin{equation}
D _{a}^{\left(cc \right)} = D _{a} - \frac{3}{2}\frac{r _{a}}{{\cal  
R}} \ \
\ \ \left(k = \pm 1\right)\ .  \label{74}
\end{equation}
Introducing here (\ref{57}) for $r _{a}$ and applying the zeroth-order
relation (\ref{32}) yields 
\begin{equation}
D _{a}^{\left(cc \right)} = \frac{2}{{\cal R}} \left[\dot{\Theta }  
D _{a} +
\Theta t _{a}\right]\ \ \ \ \left(k = \pm 1\right)\ .  \label{75}
\end{equation}
Differentiating the latter relation and using the equations of the  
system (
\ref{37}) and (\ref{38}) on its r.h.s. provides us with 
\begin{equation}
\dot{D}_{\mu }^{\left(cc \right)} = - \frac{\dot{p}}{\rho + p}D  
_{\mu } -
\Theta P _{\mu } - \frac{2}{{\cal R}}\Theta \frac{\nabla ^{2}}{a ^{2}} P
_{\mu}\ \ \ \ \left(k = \pm 1\right)\ .  \label{76}
\end{equation}
According to (\ref{36}) the first and second terms on the r.h.s. of  
(\ref{76}
) cancel and we arrive at 
\begin{equation}
\dot{D}_{\mu }^{\left(cc \right)} = - \frac{2}{{\cal R}}\Theta  
\frac{\nabla
^{2}}{a ^{2}} c _{s}^{2} D _{\mu }\ \ \ \ \ \left(k = \pm 1\right)\ .
\label{77}
\end{equation}
This simple relation comprises the entire linear perturbation  
dynamics for
nonflat background universes. In the case $k = 0$, where the  
`cc'-quantities
are not defined, one has to modify the above reasoning to arrive at a
corresponding relation. From the definition (\ref{59}) one finds  
with (\ref
{14}), 
\begin{equation}
\dot{\Theta } D _{a}^{\left(ce\right)} = \dot{\Theta } D _{a} + \Theta t
_{a} \ \ \ \ \ \ \ \left(k = 0, \pm 1 \right)\ .  \label{78}
\end{equation}
The r.h.s. of this equation coincides with the expression in the  
bracket on
the r.h.s. of (\ref{75}). After similar steps like those between  
eqs.(\ref
{75}) and (\ref{77}) one gets 
\begin{equation}
\left[a ^{2}\dot{\Theta } D _{\mu }^{\left(ce  
\right)}\right]^{\displaystyle 
\cdot} = - a ^{2}\Theta c _{s}^{2} \frac{\nabla ^{2}}{a ^{2}}D  
_{\mu } \ \ \
\ \ \ \ \left(k = 0 \pm 1\right)\ .  \label{79}
\end{equation}
Alternatively, taking into account (\ref{62}), (\ref{71}) and  
(\ref{70}),
the quantity $a ^{2}\dot{\Theta } D _{\mu }^{\left(ce \right)}$ may be
replaced to yield 
\begin{equation}
\left[a ^{2}\dot{\Theta } \frac{r _{\mu }^{\left(ce \right)}}{2 \kappa
\left(\rho + p \right)} \right]^{\displaystyle \cdot} = \left[a  
^{2}\Theta t
_{\mu }^{\left(cd \right)} \right]^{\displaystyle \cdot} = - \frac{3}{4}
\left[a ^{2} r _{\mu }^{\left(cd \right)} \right]^{\displaystyle  
\cdot} = -
a ^{2}\Theta c _{s}^{2} \frac{\nabla ^{2}}{a ^{2}}D _{\mu }\ ,   
\label{80}
\end{equation}
for $k = 0 \pm 1 $. For $k \neq 0$ equation (\ref{79}) coincides  
with (\ref
{77}). For $k = 0$ the quantities $a ^{2}r _{a}^{\left(ce \right)}$  
and $a
^{2}r _{a}^{\left(cd \right)}$ reduce to the variable $C _{a}$ used by
Dunsby and Bruni \cite{DB}.

Equation (\ref{79}) (or an equivalent formulation according to  
(\ref{80}) or
(\ref{77})) represents the most condensed form of the linear  
perturbation
dynamics. It is the main result of this paper. A comparison with  
(\ref{39})
shows that all terms in that equation, with the exception of the spatial
gradient term, have been included into a first time derivative. The  
price to
pay for this formal simplicity is that in general each of the  
equations (\ref
{77}), (\ref{79}) and (\ref{80}) couples variables that have physical
interpretations in different gauges.

It is obvious that the introduction of the new variables is especially
useful in cases where the r.h.s. of (\ref{79}) and (\ref{80}) (or  
(\ref{77}
)) may be neglected.

For $D _{\mu } = D _{\left(n \right)}\nabla _{\mu }Q _{\left(n \right)}$
(and corresponding relations for $D _{\mu }^{\left(ce \right)}$  
etc.) where
the $Q _{\left(n \right)}$ satisfy the Helmholtz equation $\nabla ^{2}Q
_{\left(n \right)} = - n ^{2} Q _{\left(n \right)}$ one has (\cite
{LiKha,Harr,KoSa,DB}) $n ^{2} = \nu ^{2}$ for $k = 0$ and $n ^{2} =  
\nu ^{2}
+ 1$ for $k = -1$, where $\nu$ is continous and related to the physical
wavelength by $\lambda = 2 \pi a/\nu$. For $k = +1$ the eigenvalue  
spectrum
is discrete, namely $n ^{2} = m \left(m + 2 \right)$ with $m =  
1,2,3....$.

On large perturbation scales, i.e., for $\nu \ll 1$ in a flat background
universe the r.h.s. of (\ref{79}) and (\ref{80}) may be neglected and we
find 
\begin{equation}
a ^{2}\dot{\Theta }D _{\left(\nu \right) }^{\left(ce \right)} = a  
^{2}\dot{
\Theta } \frac{r _{\left(\nu \right) }^{\left(ce \right)}}{2 \kappa
\left(\rho + p \right)} = a ^{2}\Theta t _{\left(\nu \right)}^{\left(cd
\right)} = - \frac{3}{4} a ^{2}r _{\left(\nu \right) }^{\left(cd  
\right)}
\approx const \mbox{\ \ \ \ \ } \left(\nu \ll 1 \right)  \label{81}
\end{equation}
for $k = 0$.

Using equation (\ref{59}) which defines $D _{a}^{\left(ce \right)}$  
in terms
of $D _{\mu }$ and $t _{\mu }$, and eliminating $t _{\mu }$ by  
(\ref{37}),
yields the following relation between $D _{a}^{\left(ce \right)}$  
and $D _{a}
$: 
\begin{equation}
\dot{\Theta }D _{\left(\nu \right) }^{\left(ce \right)} = - \Theta  
\dot{D}
_{\left(\nu \right) } + \left[\left(c _{s}^{2} - \frac{1}{3}  
\right)\Theta
^{2} - \frac{\kappa }{2}\left(\rho + 3 p \right) \right] D _{\left(\nu
\right) }\mbox{\ \ \ \ \ } \left(\nu \ll 1 \right) \ .  \label{82}
\end{equation}
Of course, the latter relation and (\ref{79}) are equivalent to  
(\ref{39}).
On large scales and for $k = 0$ equation (\ref{82}) together with  
(\ref{81})
is a first order equation to determine $D _{\left(\nu \right) }$.

In a nonflat universe (for simplicity we restrict ourselves to the  
case $k =
- 1$ which is observationally favoured) the r.h.s. of (\ref{77}),  
(\ref{79})
and (\ref{80}) do not generally vanish even for $\nu \ll 1$. The only
exception is $c _{s}^{2} \ll 1$, i.e., a dust universe  
(cf\cite{DB}). Using
the above eigenvalue structure of the Laplacian in (\ref{39}) and  
comparing
the last term in the bracket in front of $D _{\mu }$ in (\ref{39})  
with the
previous ones by taking into account the zeroth-order equation  
(\ref{31}),
one finds that the spatial gradient term may be neglected for $\nu  
\ll 1$,
provided the additional condition 
\begin{equation}
H ^{2}a ^{2} \gg 1  \label{83}
\end{equation}
with $H \equiv \Theta /3$ is satisfied. The latter relation, however,
coincides with the condition under which the curvature term in the
background equations (\ref{31}) and (\ref{32}) may be neglected.
Consequently, the quantities $D _{\mu }^{\left(cc \right)}$, $a  
^{2}\dot{
\Theta }D _{\mu }^{\left(ce \right)}$ and equivalent expressions  
under the
dot derivative on the l.h.s. of (\ref{80}) are only conserved on large
scales if the background curvature terms are negligible. Corresponding
properties for a differently defined quantity were found by Dunsby  
and Bruni 
\cite{DB}(see eq(16) in \cite{DB}. For special cases these authors,  
however,
also constructed conservation quantities without the restriction  
(\ref{83})
and on arbitrary scales.)

While the quantities $D _{\mu }^{\left(ce \right)}$, $r _{\mu  
}^{\left(ce
\right)}$, $t _{\mu }^{\left(cd \right)}$ and $r _{\mu }^{\left(cd  
\right)}$
which appear on the l.h.s. of (\ref{79}) and (\ref{80}) are well  
defined for
any value of the background curvature, the quantity $D _{\mu }^{\left(cc
\right)}$, the fractional energy density perturbation on  
hypersurfaces of
constant curvature, defined only for $k \neq 0$, appears to be the most
suitable quantity to characterize perturbations on a nonflat background,
however small the curvature terms may be. The general perturbation  
dynamics
reduces to (\ref{77}) in this case and $D _{\mu }^{\left(cc \right)}$ is
conserved on large scales if $a ^{2}H ^{2} \gg 1$.

\section{Conclusions}

We have simplified the linear, perfect fluid cosmological perturbation
theory by the introduction of new, covariant and gauge-invariant
perturbation variables. These variables are physically interpreted on
hypersurfaces of constant expansion, constant 3-curvature, or constant
energy density. Their relations to the Ellis-Bruni variables which have
physical interpretations on comoving hypersurfaces are established. The
conserved quantity in the large-scale limit for $k = 0$ may be  
expressed in
terms of the gradient of the fractional energy density on constant  
expansion
hypersurfaces or, alternatively, in terms of the expansion gradient on
constant density hypersurfaces. The most suitable quantity for the
description of the perturbation dynamics in a universe with nonvanishing
background curvature is the covariantly defined fractional, spatial  
gradient
of the energy density on hypersurfaces of constant 3-curvature. It  
is this
quantity which is approximately conserved on large scales provided the
background curvature terms are small. \newline
One may hope that use of the variables introduced in the present  
paper also
simplifies perturbation calculations under more general circumstances.

\ \newline
{\bf Acknowledgment}\newline
This paper was supported by the Deutsche Forschungsgemeinschaft.
I thank
both referees for helpful suggestions and comments.
Discussions with Diego
Pav\'{o}n and Jai-chan Hwang are gratefully acknowledged.

\ \newline

\end{document}